\documentclass[acmsmall]{acmart}
\AtBeginDocument{%
  }
\usepackage{booktabs}
\usepackage{multirow}
\usepackage{graphicx}
\usepackage{float}
\usepackage{placeins}
\usepackage{tikz}
\usepackage{pgfplots}
 \usepackage{natbib}
\setcopyright{acmlicensed}
\copyrightyear{2026}
\acmYear{2026}
\acmDOI{XXXXXXX.XXXXXXX}

\begin{document}

\title{Enhancing Large Language Models with Retrieval Augmented Generation for Software Testing and Inspection Automation}


\author{Zoe Fingleton}
\email{zfingleton@smith.edu}
\orcid{0009-0007-4184-1948}

\affiliation{
  \institution{Department of Computer Science \\
  University of Colorado Colorado Springs (UCCS)}
  \country{USA}
}

\affiliation{
  \institution{\\ Smith College, Northampton, MA}
  \country{USA}
}

\author{Nazanin Siavash}
\email{nsiavash@uccs.edu}
\orcid{0009-0000-4177-0632}
\affiliation{
  \institution{Department of Computer Science \\
  University of Colorado Colorado Springs (UCCS)}
  \country{USA}
}

\author{Armin Moin}
\email{amoin@uccs.edu}
\orcid{0000-0002-8484-7836}
\affiliation{
  \institution{Department of Computer Science \\
  University of Colorado Colorado Springs (UCCS)}
  \country{USA}
}

\renewcommand{\shortauthors}{Fingleton et al.}

\begin{abstract}
 \textbf{Abstract:} In this paper, we focus on automating two of the widely used Verification and Validation (V\&V) activities in the Software Development Lifecycle (SDLC): Software testing and software inspection (also known as review). Concerning the former, we concentrate on automated test case generation using Large Language Models (LLMs). For the latter, we enable inspection of the source code by LLMs. To address the known LLM \textit{hallucination} problem, in which LLMs confidently produce incorrect outputs, we implement a Retrieval Augmented Generation (RAG) pipeline to integrate supplementary knowledge sources and provide additional context to the LLM. Our experimental results indicate that incorporating external context via the RAG pipeline has a generally positive impact on both test case generation and code inspection. This novel approach reduces the total project cost by saving human testers'/inspectors' time. It also improves the effectiveness and efficiency of these V\&V activities, as evidenced by our experimental study. 
\end{abstract}



\maketitle

\section{Introduction}\label{sec:introduction}
The Software Development Life-Cycle (SDLC), also known as the software process, can look different across various companies and teams. However, it often comprises four fundamental stages: (i) Software Requirements Engineering (RE), (ii) Software development (which includes both software design and implementation/programming), (iii) Software Verification and Validation (V\&V), and (iv) Software deployment, evolution, and maintenance. This work is concerned with V\&V, which is a crucial stage throughout the software lifecycle. Software testing, software inspection (review), and formal verification are popular examples of V\&V activities. Formal verification is not always possible. Moreover, it is very costly. Thus, it is typically pursued for safety-critical or mission-critical systems. However, it is the only V\&V activity that can prove the absence of errors to the extent that it is verified with mathematical proofs. 

On the other extreme of the spectrum, software inspection (also known as review) resides. This is the most cost-effective V\&V activity. On average, it can find around 60\% of software defects and inefficiencies in the design and implementation, which is impressive \cite{sommerville_software_2016}. However, manual inspection is tedious and time-consuming. Large Language Models (LLMs) have \lq{}observed\rq{} many examples of software source code during their training. Therefore, they are capable of taking over the code inspection task for widely used programming languages, such as Python.

Another well-established V\&V activity is Testing. In terms of the cost, testing lies somewhere between the two extremes of formal verification and inspection. Similar to inspection, and unlike formal verification, testing can only show the presence of errors, not their absence. However, software testing is essential to ensure fidelity and adherence to the functional and non-functional requirements of software systems as specified in the RE stage of SDLC. Functional requirements are related to features and functionalities of the software system and the services that it should provide. However, non-functional requirements (also known as quality attributes) are concerned with the constraints on the software and its operation, as well as horizontal qualities of the system, such as security, privacy-preserving, reliability, adaptability, acceptability, maintainability, and sustainability.

In this work, we are interested in automated code inspection and test case generation using Artificial Intelligence (AI). Specifically, we are interested in the subdiscipline of Machine Learning, and in particular, state-of-the-art pre-trained Machine Learning models, called Large Language Models (LLMs). LLMs are fundamentally transforming the landscape of software engineering \cite{Wang+2025}. Over the past few years, researchers and practitioners have increasingly leveraged LLMs to support or automate a broad spectrum of software engineering activities, including automated test case generation, input data synthesis, test oracle construction, bug detection, debugging, and program repair. Among these, software testing has emerged as one of the most active and impactful areas of application. Given the ever-growing complexity of modern systems and the scale of testing required, organizations face significant pressure to automate testing workflows. In this context, the generative and \lq{}reasoning\rq{} abilities of LLMs offer a powerful solution, enabling more efficient handling of labor-intensive tasks.

Despite these advances, ensuring consistency and reliability in large-scale testing projects remains a considerable challenge. Software systems continuously evolve, and maintaining alignment between source code, documentation, and test artifacts is both time-consuming and error-prone \cite{RabbiSiddik2020, Yu+2017, Yichi2024}. Discrepancies between code and descriptive artifacts, such as comments or test specifications, not only reduce comprehensibility but may also introduce confusion, propagate errors, and degrade developer productivity. To mitigate such risks, the research community has proposed a variety of automated techniques over the past decade \cite{Yichi2024}. However, many of these methods struggle with the inherently informal and inconsistent nature of human-written documentation, which often includes domain-specific jargon, acronyms, inconsistent grammar, incomplete sentences, and typographical errors.

The emergence of advanced LLM architectures, exemplified by the Generative Pre-trained Transformer (GPT) series and similar models, has introduced new opportunities to overcome these limitations. Unlike traditional approaches that rely on rigid pattern matching or shallow Natural Language Processing (NLP) techniques, LLMs can analyze software repositories in a context-aware manner, capturing semantic nuances and subtle dependencies across code and documentation \cite{Yichi2024, Zhang+2025reference}. This capability enables more robust detection of inconsistencies and provides richer support for developers in maintaining coherent software artifacts throughout the lifecycle.

Rather than relying solely on manual, rule-based generation, or search-based generation strategies for test cases, which often require significant manual configuration, LLMs can dynamically synthesize test cases by reasoning about the program intent, requirements, and contextual cues extracted from code and natural language documentation. This shift represents a move toward intelligent, knowledge-driven automation, where models are not only capable of generating test cases but also refining them iteratively to improve coverage, correctness, and maintainability. By integrating LLMs into the testing pipeline, software engineering is moving closer to a future in which testing is not just automated but also adaptive, context-sensitive, and capable of evolving alongside the software itself.

To make automated code analysis and test case generation more context-aware while also improving the reliability of LLMs and reducing hallucinations, we employ a Retrieval-Augmented Generation (RAG) approach. The retrieval stage is designed to interpret the input prompt and gather supporting information from external knowledge bases \cite{Deng+2024, Wenqi+2024}. Such techniques have already proven effective in domains including content generation, question answering systems, chatbots, and conversational agents.

In NLP tasks, RAG provides additional guidance by integrating external evidence into the generation process, effectively serving as contextual reinforcement for both the input and the produced output \cite{Urvashi+2019}. Typically, the system first deploys a retriever to locate and extract the most relevant documents from the underlying database. These retrieved items are then fused with the user’s original query, creating an enriched context that steers the model toward more accurate and coherent responses \cite{GautierEdouard2020}.

To quantify the hallucination problem in LLM-based code analysis, we estimate hallucination metrics through overall accuracy. In other words, the proportion of correct outputs produced by the model serves as an indicator of how frequently the model avoids fabricating or generating misleading information. For this purpose, we experiment with multiple LLMs, each of which can be applied in different stages of the V\&V activities. Employing different LLMs allows us to capture the variation in their capabilities and ensures that our evaluation is not biased toward the strengths or weaknesses of a single model.

To rigorously validate and evaluate our proposed approach for automated code inspection, we compare the performance of the models against a human baseline. Specifically, we consider an average detection rate of human code inspectors, which has been reported to be approximately 60\%. By using this reference point, we are able to assess not only whether the LLMs can automate parts of the inspection process, but also whether they can reach or surpass the effectiveness of human experts in detecting errors and defects.

In addition, to validate and evaluate the proposed framework for automated test case generation, we adopt a reference benchmark dataset introduced by Wang et al. \cite{wang_testeval_2025}. The TestEval dataset has been specifically designed to measure the quality of automatically generated test cases across a wide range of tasks, making it a reliable choice for this study. The dataset enables us to assess the coverage of the generated test cases in a controlled and consistent setting. Moreover, by using this dataset, we can ensure that the evaluation is both reproducible and comparable with other state-of-the-art techniques. The evaluations allow us to demonstrate not only the feasibility but also the relative advantages and limitations of applying LLMs in automated code inspection and test case generation. 

The contribution of this paper is twofold: (i) We propose a novel approach to automated code inspection based on LLMs and RAGs. (ii) We propose a novel approach to automated test case generation using LLMs and RAGs. Therefore, we answer the following Research Questions (RQs) in our experimental study: RQ1. Can LLMs w/o RAG pipelines perform code inspection effectively and efficiently? RQ2. Can LLMs w/o RAG pipelines perform test case generation effectively and efficiently?

The rest of this paper is structured as follows: Section \ref{sec:background} provides some background information on LLMs and RAG. Afterward, Section \ref{sec:related-work} reviews the literature of automated V\&V. Further, we propose our novel approach in Section \ref{sec:proposed-approach}. The experimental study in Section \ref{sec:experimental-results} validates and evaluates the proposed approach. Potential threats to validity are discussed in Section \ref{sec:threats-to-val}. Finally, we conclude and suggest future work in Section \ref{sec:conclusion-future-work}.

\section{Background} \label{sec:background}

\subsection{Large Language Models (LLMs)}
LLMs are a significant advancement in AI, trained on vast corpora containing billions of words. These models are upscaled iterations of language models (LMs), which learn intricate patterns of natural language from a dataset, enabling them to predict and generate contextually relevant information \cite{zhao_survey_2023}. The driving force behind many LLMs is the self-attention mechanism found in the Transformer architecture \cite{vaswani_attention_2017}, which enables understanding of context. Through a process known as \textit{pre-training}, LLMs learn some general knowledge from a dataset. Often, this is accomplished using the aforementioned self-attention. Furthermore, beyond this ability to learn a general language understanding from initial training, LLMs can be \textit{fine-tuned} for a specific task. Fine-tuning redirects the model to focus on a specific field for enhanced domain-specific performance \cite{naveed_comprehensive_2025}. Fine-tuning gained prominence with BERT's introduction in 2019 \cite{church_emerging_2021,devlin_bert_2019}, which integrated pre-training and fine-tuning approaches. Since then, these techniques have been widely adopted in industry \cite{parthasarathy_ultimate_2024}  and scaled \cite{zhang_when_2024}.

\subsection{LLM Hallucination}
As the performance of LLMs in NLP tasks has rapidly progressed, the generation of non-factual responses has become a pressing concern. Known as \textit{hallucination}, LLMs have a tendency to produce content that is either unverifiable or contradictory to their knowledge base \cite{huang_survey_2025}. This often occurs when LLMs lack sufficient information to answer a user's query.

We can generally group hallucinations into two categories: factuality hallucinations and faithfulness hallucinations. Factuality hallucination occurs when an LLM generates a result that is factually incorrect within the broader scheme of known knowledge. In contrast, faithfulness hallucination directly contradicts an LLM's learned knowledge or user instructions \cite{huang_survey_2025}.


\subsection{Retrieval Augmented Generation (RAG)}
Introducing a context-specific knowledge source to the LLM’s accessible data is one approach to mitigating the hallucination issue. The presence of this supplementary information reduces the likelihood that the LLM will lack context \cite{bechard_reducing_2024}. To access knowledge base data, documents in the RAG knowledge base are first encoded into embeddings. The general methodology for performing these embeddings involves using a sentence transformer model or an encoder-based architecture to find sentence embeddings through contrastive learning \cite{huang_survey_2025}. 

With these documents in vector storage, the pipeline can perform a similarity search for a specific query and return relevant context. Using this methodology, an LLM can receive additional context-specific knowledge pertinent to a specific prompt. 

Existing results, particularly in the domain of NLP tasks, show that RAG-enhanced models can outperform the state-of-the-art in several language generation tasks \cite{lewis_retrieval-augmented_2021, lyu_crud-rag_2025}. Subsequently, RAG pipelines have been integrated into domain-specific contexts, such as code generation and summarization \cite{parvez_retrieval_2021}, mathematics \cite{levonian_retrieval-augmented_2023}, and medicine \cite{yang_writing_2021}, with similarly successful results \cite{zhao_retrieval-augmented_2024}. In this work, we propose deploying RAG in the V\&V activities, specifically for test case generation and source code review. 

\section{Related Work} \label{sec:related-work}





 \subsection{Code Inspection}
 
Code inspection is a crucial activity within the V\&V process; however, traditional approaches relying on manual review, where one or more developers examine their peers’ code \cite{BavotaRusso2015, Fagan1999}, are both time-consuming and labor-intensive. This limitation has motivated research into automation techniques, particularly with the rise of LLMs. In this area, a recent systematic survey of code review tasks analyzed 691 publications and identified 24 key studies published between 2015 and 2024, providing a structured overview of models, metrics, baselines, and results in automated code review \cite{HeumullerOrtmeier2025}.
Recent advances in state-of-the-art LLMs demonstrate strong performance in automated code inspection tasks \cite{guo_exploring_2024, hao_ev_2023}, especially when models are fine-tuned for domain-specific software engineering applications \cite{pornprasit_fine-tuning_2024}. Additionally, Cihan et al. \cite{Cihan+2025} highlighted the gap in longitudinal studies by examining the impact of automated code review tools in software development processes. Their research extends the scope by investigating an LLM-based automated code review tool deployed in real-world industry settings, with particular attention to its influence on development artifacts and developer perceptions.
Another line of research has explored the role of LLMs in processing code review feedback. Studies show that models can effectively interpret inspection comments and generate appropriate revisions \cite{lin_codereviewqa_2025}. In parallel, automated comment generation has gained traction, where Machine Learning and NLP methods are leveraged to analyze code and produce defect-focused annotations or fix suggestions \cite{BenSahraoui2024, Li+2022}. These approaches improve review efficiency, but remain limited in their ability to express uncertainty or identify gaps in understanding \cite{zhao_codejudge-eval_2024}. In another study, Sun et al. \cite{Sun+2025} provided the first systematic investigation of the adoption and usage of AI-based code review actions on GitHub, categorizing available tools by their review granularity (pull request, file, or hunk level) as well as by their behavior, such as the specific triggers that initiate automated actions. Building on this, an LLM-assisted framework has been introduced for evaluating the actionability of code review comments and verifying whether they have been addressed, achieving accuracy levels comparable to human annotations. 
A further challenge arises when dealing with obfuscated or intentionally complex code, where LLMs struggle to provide meaningful analyses and often exhibit degraded performance \cite{fang_large_2024}. These limitations underscore the importance of continued research into integrating LLM-based inspection tools within practical software engineering workflows, balancing their demonstrated strengths with recognition of their current shortcomings.

 \subsection{Test Case Generation}

Beyond applications for code inspection, LLMs show potential for adaptability in other areas of software development. One such area is computational test case generation, a well-established field \cite{miller_automatic_1976} that has recently begun to incorporate LLMs to facilitate complete automation. Recent research into automated unit test generation and evaluation suggests that LLMs have surpassed the current state-of-the-art in some forms of test case generation, such as JavaScript test generation programs \cite{schafer_empirical_2024}. Furthermore, LLMs demonstrate potential for automating test case generation from bug reports \cite{kang_large_2023}. 
For instance, Chen et al. \cite{Chen+2024} introduced ChatUniTest, a unit test generation framework powered by LLMs. The framework leveraged novel mechanisms such as adaptive focal context and a generation–validation–repair pipeline, culminating in the development of the ChatUniTest Core. This system was designed to support both researchers and tool builders by improving the reliability and adaptability of automated unit test generation.
Complementing this direction, Li et al. \cite{LiYuan2024} proposed TestChain, a multi-agent framework that decouples the generation of test inputs from test outputs. TestChain adopts a ReAct-style conversational chain, enabling LLMs to interact directly with a Python interpreter to produce more accurate and executable outputs. Experimental results demonstrated its effectiveness, with TestChain significantly outperforming baseline approaches. Notably, when using GPT-4 as the backbone model, TestChain achieved a 13.84\% improvement in test case accuracy over the baseline on the LeetCode-hard dataset.

Nevertheless, LLM performance remains inconsistent across datasets and code contexts, with a substantial influence from the LLMs' training sets \cite{siddiq_using_2024}. In some cases, LLMs have a similar performance to current state-of-the-art benchmarks, while in others, they are significantly outperformed. As one solution, empirical research into the performance of specific models, namely the ChatGPT suite, has shown potential for improvement upon out-of-the-box model generation ability through the implementation of iterative test refinement \cite{yuan_no_2024}.

\section{Proposed Approach} \label{sec:proposed-approach}
 \begin{figure*}[h]
\centering
\includegraphics[width=0.9\columnwidth]{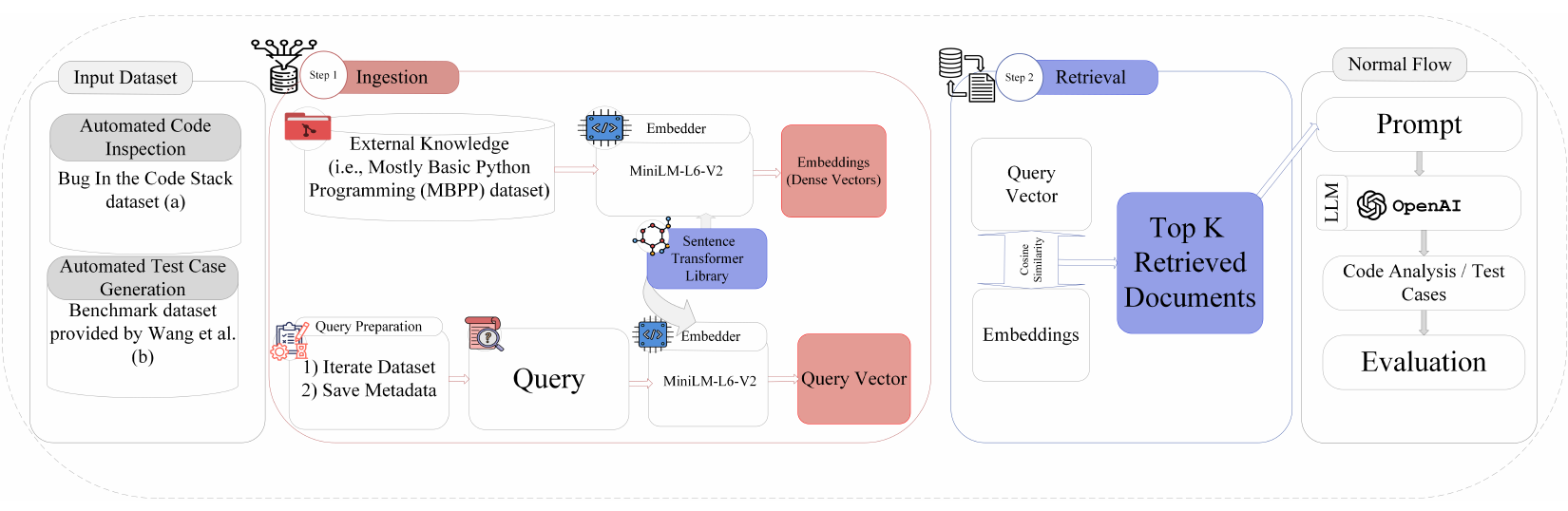}\caption{Overview of the proposed RAG-based framework for automated code inspection and test case generation. The pipeline begins with dataset ingestion, where (a) the \textit{Bug In the Code Stack} dataset from \cite{lee_bug_2024} is used for automated code inspection, and (b) the benchmark dataset from Wang et al. \cite{wang_testeval_2025} is used for automated test case generation. External knowledge (e.g., MBPP dataset \cite{austin_program_2021,google-mbpp-repo}) is embedded using MiniLM-L6-V2 via the Sentence Transformer library. In the retrieval stage, query vectors are matched with stored embeddings using cosine similarity to identify the top-K relevant documents. These retrieved contexts are then passed to the LLM for code analysis and test case generation, followed by evaluation to measure effectiveness.}
\label{fig:comp-view}
\end{figure*}

While LLMs show promise across software development, their use in V\&V remains limited due to persistent challenges with model \textit{hallucinations}, where outputs diverge from the input, context, or factual knowledge\cite{Ziyao+2020}. Such errors threaten the reliability of tasks like code inspection and automated test generation, undermining trust in LLM-driven workflows.  To address this, we propose integrating an RAG pipeline that supplies domain-specific external knowledge during generation with the goal of reducing hallucination risks. In Figure \ref{fig:comp-view}, we outline the proposed approach.

\subsection{Automated Code Inspection}
\label{sec:automated-ver}

For automated code inspection, we begin by initializing our model to work with a user-specified LLM setup. Several arguments are allowed for this configuration, including the model, temperature, max number of tokens, and whether to enable a RAG knowledge base. 

Following the processing of user arguments, we instantiate a pre-trained sentence transformer embedding model that later enables us to encode query data into embeddings. This marks the beginning of the Pipeline Setup component of the model pipeline. In our base configuration, we use all-MiniLM-L6-v2 \cite{noauthor_sentence-transformersall-minilm-l6-v2_2024} as our embedding model. Once our embedding model is configured, we implement in-memory vector storage. This will store our embeddings for easy access. With the framework for our model now established, we can load data and our knowledge documents. 

In parallel, we prepare our knowledge base corpus by loading JSON/JSONL data from a GitHub repository. As a foundational knowledge base, this pipeline uses the Mostly Basic Python Programming (MBPP) dataset developed by Austin et al. \cite{austin_program_2021, google-mbpp-repo}. This dataset contains 1000 crowd-sourced Python code snippets, useful as examples of well-written and error-free \textit{golden standards}.

Our model parses this data line by line for a task ID, content, and metadata; each task and accompanying data are saved as a separate \lq{}document.\rq{}

Following data loading and model activation, we move to the second step: knowledge base configuration (i.e., Retrieval). Here, if the RAG flag in the command-line arguments is enabled, we prepare our RAG pipeline. Using the embedding model and loaded documents from our pipeline setup, we generate embeddings for these documents and load the complete embedded corpus into the vector storage.

Following the RAG setup, we enter the main processing loop. The most focal component of this step is the query object. To create an instance of the query class, our model iterates through each data point in the loaded dataset and prepares the data as a query object. To do this, we process the data and save class variables for each relevant piece of metadata. Next, we run a function that randomly selects one of the eight code snippets. This function returns the ground truth of the snippet's validity status (bug or no bug) and the selected data point.

From here, the selected data point is encoded into an embedding, which is input into the vector store for a similarity search. To find relevant documents, we utilize the Python library NumPy to compute cosine similarity, which involves calculating the dot product of the query and document vectors and normalizing by their magnitudes, and then returning the top \textit{k} results. 

\begin{figure}[h]
    \centering
    \includegraphics[width=0.7\textwidth]{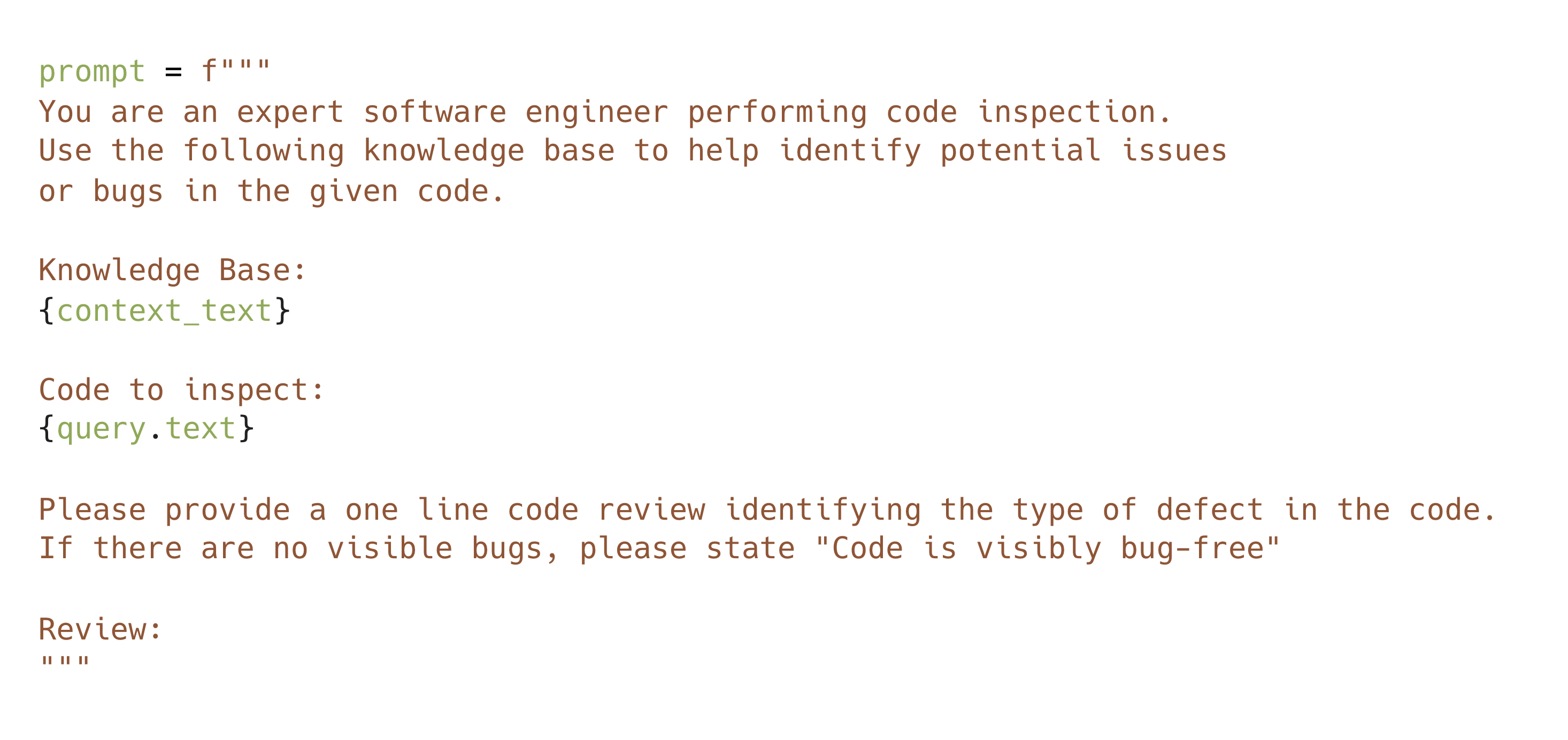}
    \caption{Prompt for code inspection}
    \label{fig_2}
\end{figure}

Using these sourced documents and the previously formatted query, the model generates a prompt with the selected output and enriches it with knowledge base context. From here, the completed prompt (see Figure \ref{fig_2}) is fed into an LLM, which then returns a code analysis of the provided snippet.

Once code analysis is produced for each data point, the data are saved and funneled into the evaluation process. With this data, we evaluate our model's code analysis results by contrasting the generated bug type with the labeled bug type collected during the selection process. If the LLM correctly identified the root of the bug (if present), its response is labeled as a \lq{}match.\rq{} Alternatively, if the LLM failed to identify a bug within the code snippet or misclassified the bug type, its response is labeled as a \lq{}mismatch.\rq{} We run this evaluation through a new LLM conversation, with OpenAI's gpt-3.5 Turbo as the default model.

To assess the effectiveness of our model's automated code inspection, we first compare the error detection rate of the RAG-enhanced and standard LLMs with the average performance benchmark for human code inspectors, which is previously established at 60\% in this paper's introduction \cite{sommerville_software_2016}. Then, we compare performance with and without the presence of a RAG knowledge base.

\subsection{Automated Test Case Generation}
\label{sec:automated-val}

For the pipeline and knowledge base setup of the automated test case generation, we prepare our model using the same methodology as automated code inspection, creating and configuring our embedding model and vector storage. 

In our main processing loop, we use the same general framework as automated code inspection, with some systemic differences. To prepare our queries after processing each data point as a query class object, we save only the program code, function name, and function description. From here, we embed the query and use the same similarity search functionality previously described.

\begin{figure}[h]
    \centering
    \includegraphics[width=0.7\textwidth]{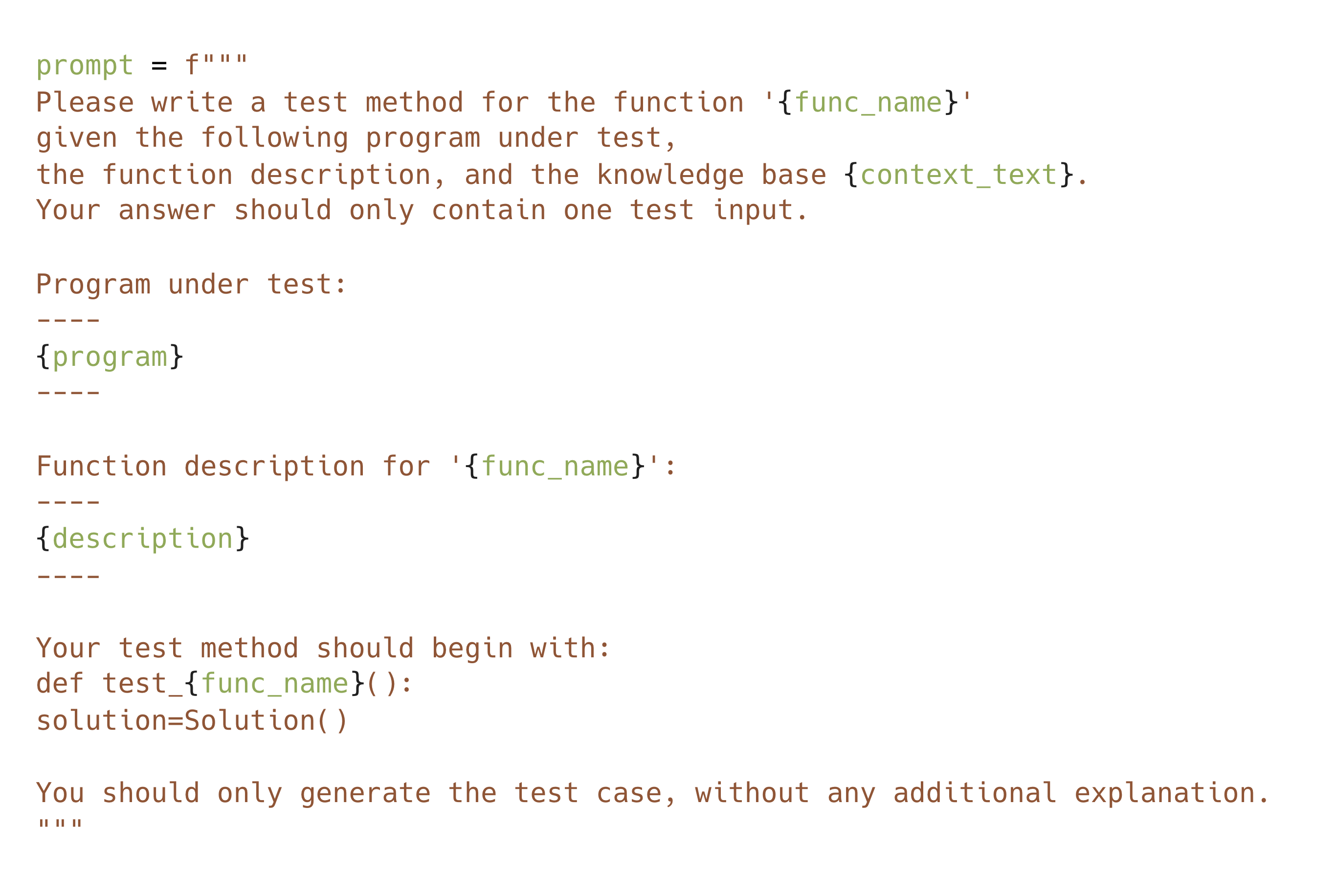}
    \caption{Prompt for test generation}
    \label{fig_3}
\end{figure}

Following embedding, the pipeline provides the specified LLM with the program under test, the function name and description, and the knowledge base context (see Figure \ref{fig_3}). The completed prompt is fed into an LLM, which then returns \textit{n} unique test cases.

To ensure that each test case is independent of previous generations, we first run the query and document context through a single test generation function. Following the completion of this cycle, the result is fed into a multi-round generation function that takes all previous generations as input.

Once raw test cases are produced for each data point, the data are saved and funneled into evaluation. We pre-process our data by formatting the results into a structured form. With this data, we evaluate our model's test case generation results in terms of overall, line, path, and branch coverage. To achieve this, we utilize TestEval's source code, which cross-references generated test cases with a complementary metadata category, \lq{}branches,\rq{} that details all the task's branches. Coverage is determined by the proportion of lines/branches covered by at least one test case using pytest-cov.

\begin{table*}[htbp]
\centering
\caption{Results of automated test generation on the TestEval dataset (210 tasks) \cite{noauthor_llm4softwaretestingtesteval_2025}}
\label{tab:coverage_results}
\resizebox{\textwidth}{!}{%
\begin{tabular}{@{}llcccccccccccc@{}}
\toprule
\multirow{2}{*}{Model} & \multirow{2}{*}{RAG} & \multicolumn{2}{c}{Compilation success} & \multicolumn{2}{c}{Test case coverage} & \multicolumn{3}{c}{Line $cov@k$} & \multicolumn{3}{c}{Branch $cov@k$} \\
\cmidrule(lr){3-4} \cmidrule(lr){5-6} \cmidrule(lr){7-9} \cmidrule(lr){10-12}
& & syntax & execution & line & branch & $k=1$ & $k=2$ & $k=5$ & $k=1$ & $k=2$ & $k=5$ \\
\midrule
GPT-3.5 Turbo & X & 99.95 & 94.61 & 93.26 & 89.96 & 85.66 & 87.47 & 89.76 & 78.05 & 80.80 & 84.43 \\
GPT-3.5 Turbo & \checkmark & \textbf{100} & \textbf{98.76} & \textbf{96.64} & \textbf{93.20} & \textbf{88.35} & \textbf{90.03} & \textbf{92.63} & \textbf{80.35} & \textbf{83.02} & \textbf{87.10} \\
\midrule
GPT-4o & X & \textbf{99.90} & 98.80 & 98.27 & 96.84 & 89.00 & 91.13 & 93.66 & 81.56 & 84.90 & 89.03 \\
GPT-4o & \checkmark & 99.80 & \textbf{99.33} & \textbf{98.51} & \textbf{96.85} & \textbf{89.39} & \textbf{91.45} & \textbf{94.02} & \textbf{81.89} & \textbf{85.12} & \textbf{89.19} \\
\bottomrule
\end{tabular}%
}
\end{table*}

\begin{table}[htbp]
\centering
\caption{Efficiency of automated test generation on the TestEval dataset (210 tasks) \cite{noauthor_llm4softwaretestingtesteval_2025}}
\label{tab:test_gen_eff}
{\begin{tabular}{@{}llcc@{}}
\toprule
Model & RAG & \multicolumn{1}{c}{Runtime} \\
\midrule
GPT-3.5 Turbo & X & 1:32:30 \\
GPT-3.5 Turbo & \checkmark & \textbf{1:28:59} \\
\midrule
GPT-4o & X & \textbf{1:55:26} \\
GPT-4o & \checkmark & 2:09:43 \\
\bottomrule
\end{tabular}%
}
\end{table}

\section{Experimental Results} \label{sec:experimental-results}

\subsection{Experimental Dataset}

For our automated code inspection experimental data, we use the Bug In the Code Stack dataset provided by Lee et al \cite{lee_bug_2024}. This dataset comprises 4,010 programming tasks. For each task in this dataset, there are eight code snippets: a bug-free code sample and seven \textit{buggy} versions (e.g., the correct code missing a colon or misusing a keyword as a variable name). The dataset also contains metadata specifying the line at which buggy code occurs. 

For our automated test case generation dataset, we locally use the benchmark dataset provided by Wang et al \cite{wang_testeval_2025}. This repository's data comprises coding assessments from LeetCode, an online programming platform, and contains code for computing line and branch coverage. 

As the central knowledge base, both pipelines will be using the Mostly Basic Python Programming (MBPP) dataset seen in the code inspection model architecture \cite{austin_program_2021,google-mbpp-repo}. This dataset is relevant to automated inspection, as it contains 1000 examples of well-written code, and is relevant to automated test case generation, as it contains corresponding sets of Python problems, solutions, and test cases.

\subsection{Experimental Setup}

In this experimental study, we evaluate three commercial LLMs from the OpenAI GPT line (GPT-3.5 Turbo, GPT-4o, and GPT-4.1) using the Chat Completions API. All experiments were conducted between May and Aug 2025 to ensure consistency in model availability. To minimize stochastic variation, we fixed the temperature at 0, yielding deterministic outputs, and constrained responses to a maximum of 256 tokens, which provides concise yet sufficiently informative outputs. 

Prompts were designed in a zero-shot format and kept identical across models to ensure fair comparison. Each prompt was executed programmatically through a Python-based evaluation framework to guarantee reproducibility and consistent logging. 

\subsection{Experimental Analysis}
\subsubsection{Automated Test Case Generation}
For automated test case generation, we test the coverage of LLM-provided test cases. To evaluate automated test case generation ability, we directly implement a RAG pipeline alongside TestEval's source code. Using command line arguments as a RAG \lq{}toggle\rq{}, we can compare performance on the same data with and without a supplementary knowledge base. With this toggle determining whether a RAG database will be provided, we prompt our model to generate \textit{N} unique test cases. In this experimental configuration, \textit{N} = 20. 

Following the generation of the provided quantity of test cases, we evaluate the model's performance by computing compilation success, in terms of syntax and execution, test case coverage, in terms of line and branch coverage, and a proprietary coverage metric developed by Wang et al. in TestEval, \textit{cov@k}.

All evaluation methodology for test case generation remains unchanged from TestEval's source code, as of July 2025. To compute compilation success, the program attempts to execute without syntax errors. For coverage metrics, pytest-cov, a plugin that produces coverage reports, is used to determine the proportion of
branches covered by at least one test case. Finally, \textit{cov@k} randomly samples \textit{k} generated test cases and finds accuracy. Our findings for test case coverage are outlined in Table \ref{tab:coverage_results}. 

\begin{figure}[h]
    \centering
    \includegraphics[width=0.5\textwidth]{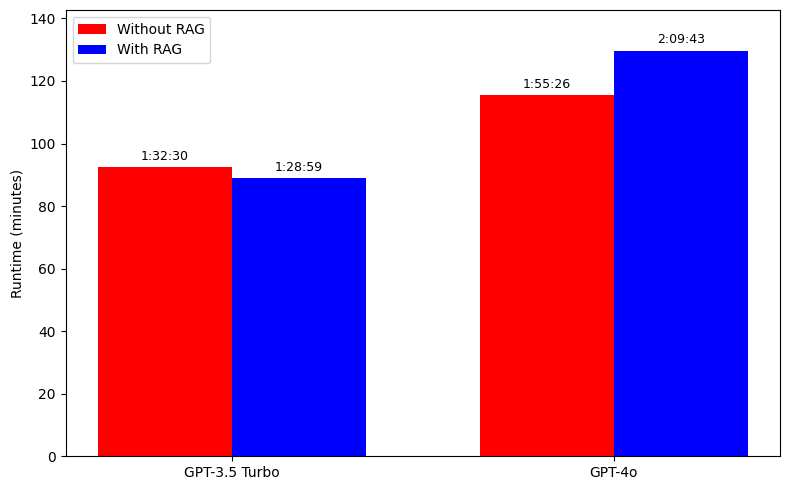}
    \caption{Efficiency of Automated Test Generation}
    \label{automated-code-generation}
\end{figure}

Regarding these results, several observable trends are apparent. Firstly, under the current experimental configuration, RAG-enhanced LLMs consistently outperform standard models in terms of execution success. Furthermore, RAG-enhanced LLMs have a greater rate of coverage across all areas of measured line and branch coverage, including \textit{cov@k}. Compilation success in terms of syntax appears to be relatively uninfluenced by the presence of a RAG pipeline. These results strongly indicate that the addition of a RAG pipeline has an overall positive influence on automated test case generation, with no substantial drawbacks visible at this time.
According to Table \ref{tab:test_gen_eff} and also the visualization in Figure \ref{automated-code-generation} , for automated test generation on the TestEval dataset, GPT-3.5 Turbo shows a clear runtime benefit from RAG, completing tasks about 3.5 minutes faster (1:32:30 → 1:28:59), suggesting retrieval helped streamline generation. Conversely, GPT-4o’s runtime increases by over 14 minutes with RAG (1:55:26 → 2:09:43), indicating that retrieval overhead outweighed any efficiency gains. Overall, GPT-3.5 Turbo with RAG is optimal for speed in this setting, while GPT-4o performs best without RAG.
\subsubsection{Automated Code Inspection}
\begin{figure}[h]
\centering
\includegraphics[width=1\columnwidth]{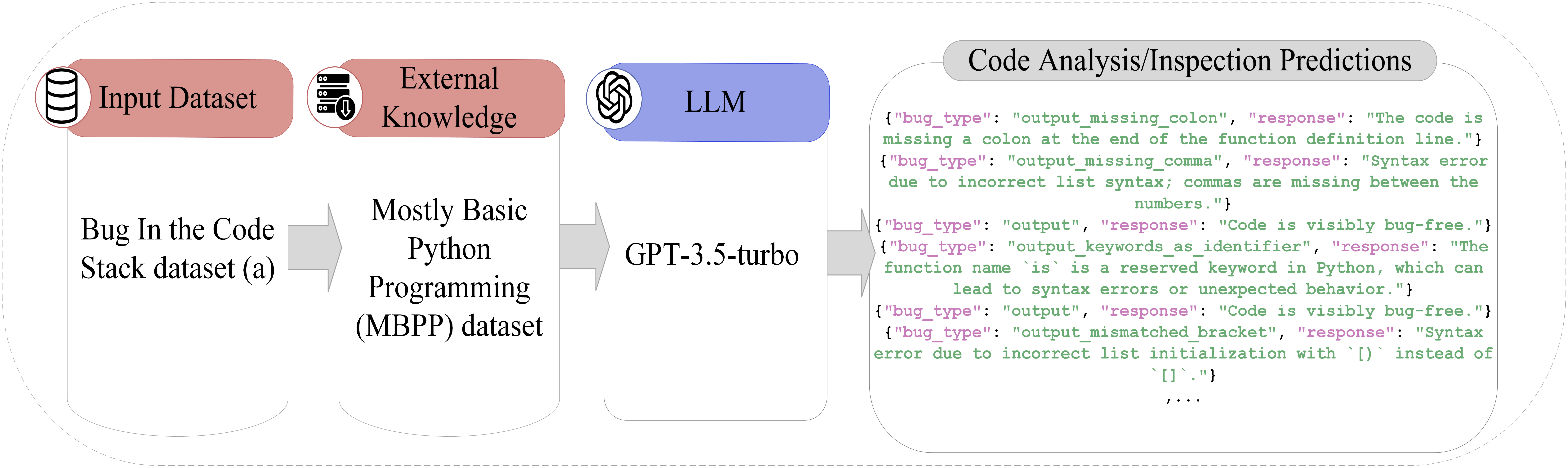}\caption{Code Analysis Predictions using GPT-3.5-turbo, where (a) the \textit{Bug In the Code Stack} dataset from \cite{lee_bug_2024} is used as an input dataset}
\label{fig:code-analysis-prediction}
\end{figure}

For automated code inspection, we test LLM accuracy in correctly identifying bugs in code snippets. Using the same configurable RAG pipeline outlined in automated test case generation, we prompt an LLM to locate and classify code defects in a provided snippet of code. For our evaluation, we compare the LLM-assigned bug label with the ground truth sourced from the Bug In the Code Stack dataset. This comparison is performed using OpenAI's GPT-3.5 Turbo \cite{openai_gpt-35_2024}. Our findings are outlined in Tables \ref{tab:review_results} and \ref{tab:mismatch_results_pro}, and Figure \ref{fig:all_mismatch_comparison1}, \ref{fig:all_mismatch_comparison2}, \ref{fig:all_mismatch_comparison3}, and \ref{fig:all_mismatch_comparison}. Also, Figure \ref{fig:code-analysis-prediction} illustrates the results of code analysis performed by GPT-3.5-Turbo with external context enabled through the RAG pipeline. The model identifies multiple bug types, including missing colons in function definitions, missing commas in list syntax, use of reserved keywords as identifiers, and mismatched brackets. In some cases, the model correctly reports that the code is free of errors.

When comparing the performance of standard models to those enhanced with RAG, several notable patterns emerge. First, in Table \ref{tab:review_results}, we observe that all models show improved accuracy in bug detection when they have access to a RAG database.

In Figures \ref{fig:all_mismatch_comparison1}, \ref{fig:all_mismatch_comparison2}, \ref{fig:all_mismatch_comparison3} , and \ref{fig:all_mismatch_comparison}, we can evaluate model performance in classifying specific bug types. RAG-enhanced LLMs consistently perform better in identifying \lq{}missing comma\rq{}, \lq{}mismatched quotation\rq{}, \lq{}mismatched bracket\rq{}, and \lq{}keyword as identifier\rq{} type bugs. The \lq{}bug-free code\rq{}, \lq{}missing colon\rq{}, \lq{}missing parenthesis\rq{}, and \lq{}missing quotation\rq{} bug types are more variable, although RAG-enhanced LLMs have a better overall performance.

Table \ref{tab:mismatch_results_pro} shows overall trends in which bug types are more challenging for the LLMs to classify. There are fewer consistent patterns present in these data, although we do see a uniformly stronger performance by RAG-enhanced models in identifying \lq{}mismatched quotation\rq{}, \lq{}mismatched bracket\rq{}, and \lq{}keyword as identifier\rq{} types of bugs. Consistent with our findings at this point, the \lq{}bug-free code\rq{}, \lq{}missing colon\rq{}, and \lq{}missing parenthesis\rq{} bug-types are less evenly distributed, as well as the \lq{}missing comma\rq{} label. However, again, we see that RAG-enhanced LLMs have a better overall performance. Our results show that incorporating a RAG pipeline consistently enhances the accuracy of automated code inspection. Furthermore, RAG-enhanced LLMs frequently outperform standard LLMs in code inspection across bug types.

\begin{figure}[h]
    \centering
    \includegraphics[width=0.5\textwidth]{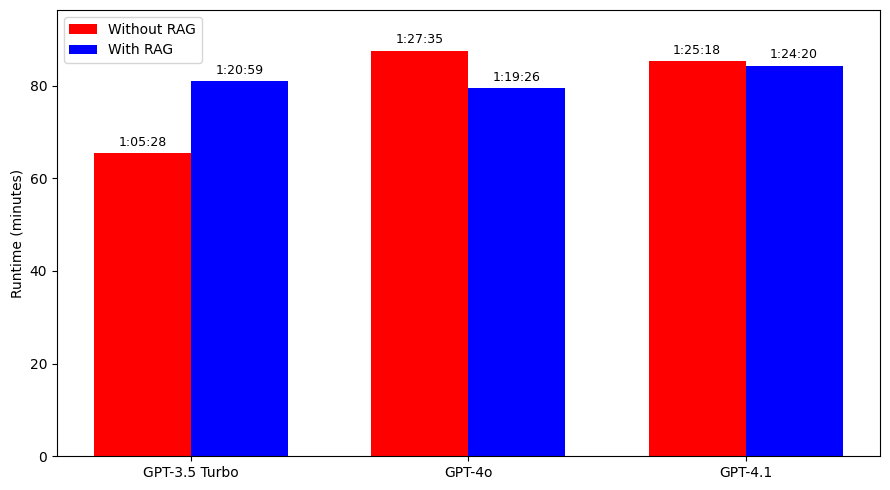}
    \caption{Efficiency of Automated Code Inspection}
    \label{automated-code-inspection}
\end{figure}

Also, the results in Table \ref{tab:review_eff} and also the visualization in Figure \ref{automated-code-inspection} show that GPT-3.5 Turbo without RAG achieves the fastest runtime (1:05:28), while adding RAG increased its runtime by about 15 minutes due to retrieval overhead. In contrast, GPT-4o benefits the most from RAG, reducing runtime by roughly 8 minutes (1:27:35 → 1:19:26), suggesting that retrieval helped streamline its code inspection process. GPT-4.1 sees only a marginal improvement of about 1 minute with RAG, indicating limited runtime sensitivity to retrieval. Overall, GPT-3.5 Turbo without RAG is optimal for speed, GPT-4o with RAG offers the best runtime improvement, and GPT-4.1’s runtime performance is largely unaffected by retrieval.

\begin{figure}[h]
\centering
\includegraphics[width=0.9\columnwidth]{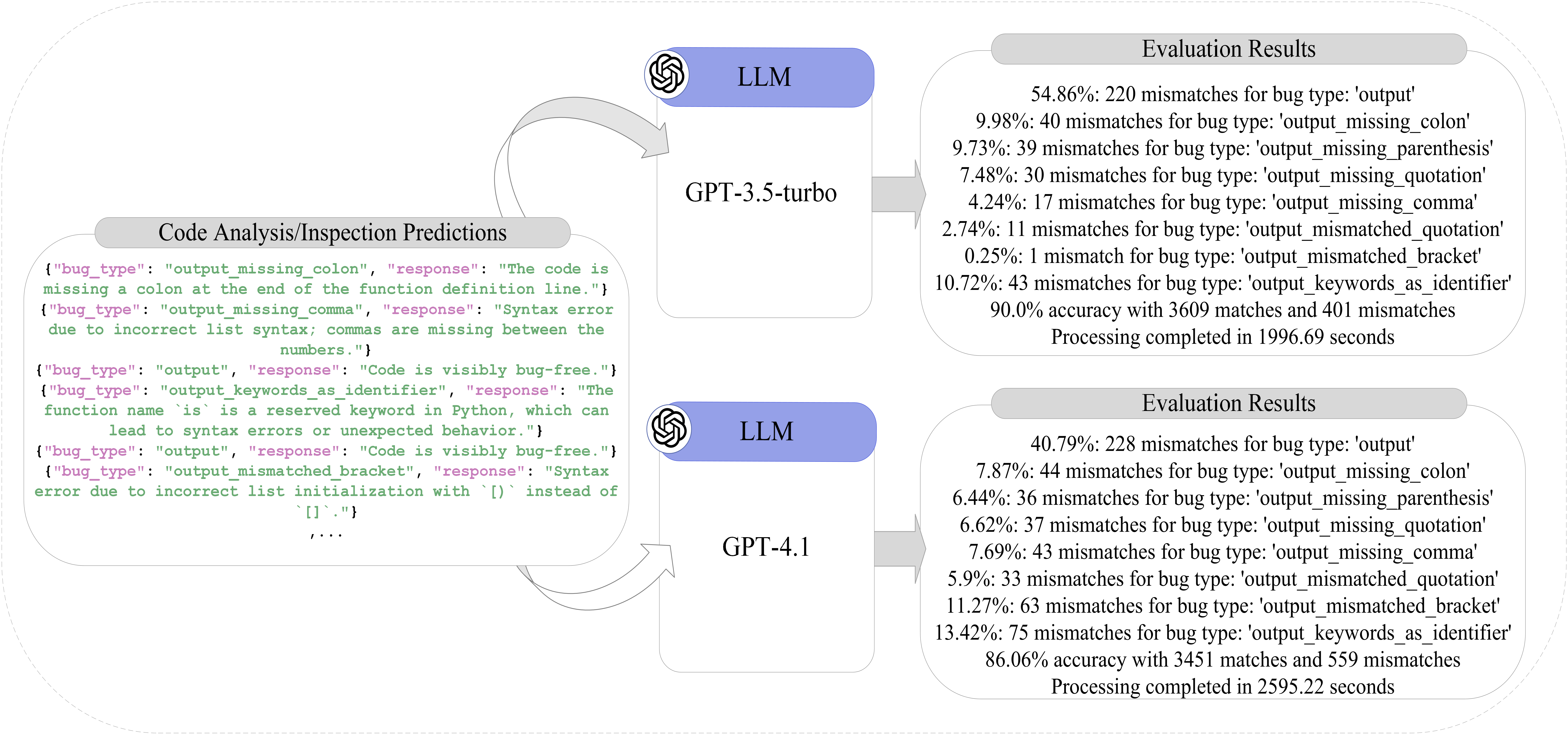}\caption{Evaluation Results of Code Analysis/Inspection Predictions}
\label{fig:review-eval}
\end{figure}

The evaluation results of the code analysis using GPT-3.5-Turbo and GPT-4.1 are presented in Figure \ref{fig:review-eval}. 

Note that the results in Figures \ref{fig:code-analysis-prediction} and \ref{fig:review-eval} come from a separate run compared to the results in the tables. The difference in outcomes is due to the inherent variability of the LLM, which can occur even with the temperature set to 0.

\begin{table*}[htbp]
\centering
\caption{Mismatch type rate (proportional to overall mismatches)}
\label{tab:mismatch_results_pro}
\resizebox{\textwidth}{!}{%
\begin{tabular}{@{}llcccccccc@{}}
\toprule
\multirow{2}{*}{Model} & 
\multirow{2}{*}{RAG} & 
\multirow{2}{*}{Bug-free code} & 
\multirow{2}{*}{Missing colon} &
\multirow{2}{*}{Missing parenthesis} &
\multirow{2}{*}{Missing quotation} & 
\multirow{2}{*}{Missing comma} &
\multirow{2}{*}{Mismatched quotation} &
\multirow{2}{*}{Mismatched bracket} & 
\multirow{2}{*}{Keyword as identifier} \\ \\
\midrule
GPT-3.5 Turbo & X & \textbf{2.55\%} & 27.03\% & 13.74\% & \textbf{5.78\%} & 7.13\% & 5.41\% & 6.53\% & 31.83\% \\
GPT-3.5 Turbo & \checkmark & 47.35\% & \textbf{8.47\%} & \textbf{8.2\%} & 10.32\% & \textbf{6.88\%} & \textbf{4.5\%} & \textbf{0.53\%} &\textbf{ 13.76\%} \\
\midrule
GPT-4 & X & \textbf{24.46\%} & 28.3\% & \textbf{2.62\%} & \textbf{5.9\%} &\textbf{ 3.75\%} & 5.06\% & 4.22\% & 25.68\% \\
GPT-4 & \checkmark & 50.0\% & \textbf{8.96\%} & 8.46\% & 7.46\% & 6.97\% & \textbf{3.48\%} & \textbf{1.0\%} & \textbf{13.68\%} \\
\midrule
GPT-4o & X & 48.55\% & \textbf{4.34\%} & 7.07\% & \textbf{6.75} & 5.47\% & 4.18\% & 2.09\% & 21.54\%\\
GPT-4o & \checkmark & \textbf{48.52\%} & 10.59\% & \textbf{6.4\%} & 12.07\% & \textbf{4.43\%} & \textbf{3.94\%} & \textbf{0.74\%} & \textbf{13.3\%}\\
\bottomrule
\end{tabular}%
}
\end{table*}
\begin{figure*}[htbp]
\centering

\caption{GPT-3.5 Turbo: Mismatch Type Count Comparison (RAG vs Non-RAG)}
\label{fig:all_mismatch_comparison1}
\begin{tikzpicture}
\begin{axis}[
    ybar,
    width=\textwidth,
    height=4cm,
    symbolic x coords={
        Bug-free code,
        Missing colon,
        Missing parenthesis,
        Missing quotation,
        Missing comma,
        Mismatched quotation,
        Mismatched bracket,
        Keyword as identifier
    },
    xtick=data,
    x tick label style={rotate=45, anchor=east, font=\small},
    ylabel={Count},
    legend pos=north west,
    ymin=0,
    ymax=550,
    bar width=6pt,
    enlarge x limits=0.15,
    ylabel near ticks,
]
\addplot[fill=red!70, draw=red] coordinates {
    (Bug-free code, 34)
    (Missing colon, 360)
    (Missing parenthesis, 183)
    (Missing quotation, 77)
    (Missing comma, 95)
    (Mismatched quotation, 72)
    (Mismatched bracket, 87)
    (Keyword as identifier, 424)
};
\addplot[fill=blue!70, draw=blue] coordinates {
    (Bug-free code, 179)
    (Missing colon, 32)
    (Missing parenthesis, 31)
    (Missing quotation, 39)
    (Missing comma, 26)
    (Mismatched quotation, 17)
    (Mismatched bracket, 2)
    (Keyword as identifier, 52)
};
\legend{Without RAG, With RAG}
\end{axis}
\end{tikzpicture}

\vspace{0.5cm}

\caption{GPT-4: Mismatch Type Count Comparison (RAG vs Non-RAG)}
\label{fig:all_mismatch_comparison2}
\begin{tikzpicture}
\begin{axis}[
    ybar,
    width=\textwidth,
    height=4cm,
    symbolic x coords={
        Bug-free code,
        Missing colon,
        Missing parenthesis,
        Missing quotation,
        Missing comma,
        Mismatched quotation,
        Mismatched bracket,
        Keyword as identifier
    },
    xtick=data,
    x tick label style={rotate=45, anchor=east, font=\small},
    ylabel={Count},
    legend pos=north west,
    ymin=0,
    ymax=450,
    bar width=6pt,
    enlarge x limits=0.15,
    ylabel near ticks,
]
\addplot[fill=red!70, draw=red] coordinates {
    (Bug-free code, 261)
    (Missing colon, 302)
    (Missing parenthesis, 28)
    (Missing quotation, 63)
    (Missing comma, 40)
    (Mismatched quotation, 54)
    (Mismatched bracket, 45)
    (Keyword as identifier, 274)
};
\addplot[fill=blue!70, draw=blue] coordinates {
    (Bug-free code, 201)
    (Missing colon, 36)
    (Missing parenthesis, 34)
    (Missing quotation, 30)
    (Missing comma, 28)
    (Mismatched quotation, 14)
    (Mismatched bracket, 4)
    (Keyword as identifier, 55)
};
\legend{Without RAG, With RAG}
\end{axis}
\end{tikzpicture}

\vspace{0.5cm}

\caption{GPT-4o: Mismatch Type Count Comparison (RAG vs Non-RAG)}
\label{fig:all_mismatch_comparison3}
\begin{tikzpicture}
\begin{axis}[
    ybar,
    width=\textwidth,
    height=4cm,
    symbolic x coords={
        Bug-free code,
        Missing colon,
        Missing parenthesis,
        Missing quotation,
        Missing comma,
        Mismatched quotation,
        Mismatched bracket,
        Keyword as identifier
    },
    xtick=data,
    x tick label style={rotate=45, anchor=east, font=\small},
    ylabel={Count},
    legend pos=north west,
    ymin=0,
    ymax=450,
    bar width=6pt,
    enlarge x limits=0.15,
    ylabel near ticks,
]
\addplot[fill=red!70, draw=red] coordinates {
    (Bug-free code, 302)
    (Missing colon, 27)
    (Missing parenthesis, 44)
    (Missing quotation, 42)
    (Missing comma, 34)
    (Mismatched quotation, 26)
    (Mismatched bracket, 13)
    (Keyword as identifier, 134)
};
\addplot[fill=blue!70, draw=blue] coordinates {
    (Bug-free code, 197)
    (Missing colon, 43)
    (Missing parenthesis, 26)
    (Missing quotation, 49)
    (Missing comma, 18)
    (Mismatched quotation, 16)
    (Mismatched bracket, 3)
    (Keyword as identifier, 54)
};
\legend{Without RAG, With RAG}
\end{axis}
\end{tikzpicture}

\caption{Mismatch Type Count Comparison (GPT-3.5, GPT-4, GPT-4o) between RAG and Non-RAG}
\label{fig:all_mismatch_comparison}
\end{figure*}

\begin{table}[htbp]
\centering
\caption{Automated code inspection results on the Bug In the Code Stack dataset (4,010 tasks) \cite{noauthor_hamminghqbug---code-stack_2024}}
\label{tab:review_results}
\resizebox{0.4\textwidth}{!}{%
\begin{tabular}{@{}llcccccc@{}}
\toprule
\multirow{2}{*}{Model} &  
\multirow{2}{*}{RAG} & 
\multirow{2}{*}{Matches} & 
\multirow{2}{*}{Mismatches} & 
\multirow{2}{*}{Accuracy} \\ \\
\midrule
GPT-3.5 Turbo & X & 2678 & 1332 & 66.78 \\
GPT-3.5 Turbo & \checkmark & \textbf{3632} & \textbf{378} & \textbf{90.57} \\
\midrule
GPT-4 & X & 2943 & 1067 & 73.39 \\
GPT-4 & \checkmark & \textbf{3608} & \textbf{402} & \textbf{89.98}\\
\midrule
GPT-4o & X & 3388 & 622 & 84.49 \\
GPT-4o & \checkmark & \textbf{3604} & \textbf{406} & \textbf{89.88}\\
\bottomrule
\end{tabular}%
}
\end{table}


\begin{table}[htbp]
\centering
\caption{Efficiency of automated code inspection on the Bug In the Code Stack dataset (4,010 tasks) \cite{noauthor_hamminghqbug---code-stack_2024}}
\label{tab:review_eff}
{\begin{tabular}{@{}llcc@{}}
\toprule
Model & RAG & \multicolumn{1}{c}{Runtime} \\
\midrule
GPT-3.5 Turbo & X & \textbf{1:05:28} \\
GPT-3.5 Turbo & \checkmark & 1:20:59 \\
\midrule
GPT-4o & X & 1:27:35 \\
GPT-4o & \checkmark & \textbf{1:19:26} \\
\midrule
GPT-4.1 & X & 1:25:18 \\
GPT-4.1 & \checkmark & \textbf{1:24:20} \\
\bottomrule
\end{tabular}%
}
\end{table}

\section{Threats to Validity}
\label{sec:threats-to-val}
While our proposed approach aims to mitigate hallucinations in LLM-based V\&V, several validity threats remain. In the following, we discuss threats to Internal, External, and Construct Validity.
\subsection{Internal Validity}
Our evaluation depends on specific datasets (e.g., MBPP) and selected LLM configurations. Biases or limitations in these datasets (such as incomplete coverage of real-world code defects or limited test case diversity) may have influenced the observed effectiveness of the RAG-enhanced pipeline. Moreover, hyperparameter settings (e.g., temperature, context length, retrieval top-\textit{k}) could affect model behavior in ways not fully controlled in our experiments. As part of our mitigation strategy, in future work, we will expand to additional datasets, incorporate diverse benchmark suites, and perform hyperparameter tuning. We also plan to adopt adaptive retrieval strategies that dynamically adjust retrieval parameters to reduce configuration-induced variance.
\subsection{External Validity}
The generalizability of our findings is constrained by the domains and datasets used. While MBPP and TestEval are widely adopted in prior studies, they may not fully capture the complexity of large-scale industrial software systems. As such, our results may not directly transfer to highly domain-specific or safety-critical environments (e.g., healthcare or finance), where codebases, defect types, and testing requirements differ significantly. Future studies will evaluate the pipeline on large, real-world, domain-specific repositories and integrate additional external knowledge sources as context.
\subsection{Construct Validity}
The metrics used to measure effectiveness, such as error detection rate for inspection and coverage metrics (line, branch, path) for test generation, provide useful proxies for reliability but do not fully capture the broader notion of software quality. For instance, high test coverage does not guarantee defect detection, and identifying a bug type may not equate to producing actionable fixes. Additionally, hallucination reduction was assessed indirectly through improvements in factual correctness and alignment with ground truth, which may not capture all forms of unfaithfulness. To address these threats, we plan to incorporate semantic-oriented metrics such as CodeBLEU, extend the evaluation to include fix viability and actionable repair quality, and adopt finer-grained hallucination assessments.
\subsection{Conclusion Validity}
Our comparative analysis between RAG-enhanced and baseline LLMs is based on limited experimental runs and a single human performance benchmark (60\%  accuracy). Statistical fluctuations or variations in LLM outputs may affect reproducibility. Furthermore, conclusions regarding efficiency and effectiveness gains must be interpreted cautiously, as retrieval introduces computational overhead that may scale differently in larger deployments. To mitigate these threats, future work will include repeated trials, cross-run variance analysis, and statistical significance testing to ensure robust conclusions. We will also profile retrieval overhead more extensively and examine scalability under larger workloads and alternative deployment configurations.



\section{Conclusion and Future Work} \label{sec:conclusion-future-work}
In this paper, we have proposed a novel approach to automated software test case generation and automated code inspection using LLMs. Our experimental results indicate that LLMs hold significant potential for applications in software testing and inspection. Regarding our initial research questions, \lq{}Can LLMs w/o RAG perform code inspection effectively and efficiently?\rq{} and \lq{}Can LLMs w/o RAG perform test case generation effectively and efficiently?\rq{}, we have validated these questions. In particular, we have found that the addition of a RAG pipeline generally increases the effectiveness of these models. Notably, RAG-enhanced models consistently outperform their standard counterparts in terms of test case generation coverage and their overall accuracy in identifying and classifying bugs during code inspection.

Future studies will investigate adaptive retrieval strategies that trigger retrieval only when they are likely to be beneficial, thereby reducing unnecessary overhead. Scaling the experiments to a broader range of models, especially more recent ones such as GPT-5, which was not available at the time of our evaluation, would help confirm the extent to which the observed runtime and performance patterns generalize better. Detailed profiling of the retrieval pipeline may also reveal optimization opportunities, while exploring hybrid frameworks that combine selective retrieval with in-context learning could offer a better balance between speed and accuracy. In addition, future work will evaluate alternative external knowledge sources to understand how different context types influence code generation quality and efficiency. Complementary metrics, such as CodeBLEU, will be used to assess semantic correctness, providing a more holistic evaluation. Finally, we plan to incorporate statistical significance testing to determine whether the observed differences are robust and not due to random variation.

\section*{Data Availability}
The source code and data are accessible at \cite{qaslabrepo}. This repository contains code to run and evaluate automated test case generation and automated code review using OpenAI, Ollama, and Hugging Face models. Additionally, this work uses data from Google's Mostly Basic Python Programming dataset \cite{google-mbpp-repo}, the Bug In the Code Stack dataset by Lee et al. \cite{lee_bug_2024}, and the TestEval dataset from Wang et al. \cite{wang_testeval_2025}. The relevant experimental data used in this study are accessible in the source code repository. To reproduce the experimental setup, users will need to download TestEval's code \cite{noauthor_llm4softwaretestingtesteval_2025} from the source. Details are specified in the README.md for this paper's repository.

\section*{Acknowledgment}
This material is based upon work supported by the U.S. National Science Foundation (NSF) under Grant No. 2349452. Any opinions, findings, conclusions, or recommendations expressed in this material are those of the authors and do not necessarily reflect the views of the NSF.
In preparing this work, we used generative AI models and tools, including the OpenAI GPT models, to assist with generating and revising content, including code and text.
Finally, the authors would like to thank Himon Thakur for discussions and comments.
\bibliographystyle{ieeetr}
\bibliography{refs}


\end{document}